\documentclass{aa}  

\usepackage{mwe}
\usepackage{graphicx}
\usepackage{subfig}
\usepackage{float}
\usepackage{multicol}
\usepackage{lipsum}
\usepackage{ragged2e}
\usepackage{eurosym}
\usepackage{multirow}
\usepackage{indentfirst}
\usepackage[table]{xcolor}
\usepackage{graphics}
\usepackage{adjustbox}
\usepackage{hyperref} 
\usepackage{natbib}

\hypersetup{
    colorlinks=true,
    linkcolor=blue,
    filecolor=magenta,
    urlcolor=cyan,
    pdftitle={Orbital dynamics in galactic potentials under mass transfer},
    pdfpagemode=FullScreen,
    }
    
\begin{document}

   \title{Orbital dynamics in galactic potentials under mass transfer}

   \author{Eduárd Illés 
          \inst{1,2}
          \and
          Dániel Jánosi \inst{2,3}
          \and
          Tamás Kovács\inst{4,5}%\fnmsep\thanks{Just to show the usage
          %of the elements in the author field}
          }

   \institute{Lund University, Division of Astrophysics, Department of Physics, Box 43, Lund SE-221 00, Sweden
        \and
            Department of Theoretical Physics, Eötvös Loránd University, Pázmány Péter sétány 1A, H-1117 Budapest, Hungary \\
         \email{eduard.illes.8687@student.lu.se}
        \and
            HUN-REN Institute of Earth Physics and Space Science, Csatkai Endre utca 6-8, H-9400 Sopron, Hungary \\
        %\and
        %    Institute of Nuclear Techniques, Budapest University of Technology and Economics, Műegyetem rakpart 3, H-1111 Budapest, Hungary \\
        \email{daniel.janosi@ttk.elte.hu}
        \and 
            Department of Atomic Physics, Eötvös Loránd University, Pázmány Péter sétány 1A, H-1117 Budapest, Hungary
         \and
             HUN-REN - ELTE Extragalactic Astrophysics Research Group, Pázmány Péter sétány 1A, H-1117, Budapest, Hungary\\
             \email{tamas.kovacs@ttk.elte.hu}
             %\thanks{The university of heaven temporarily does not
             %        accept e-mails}
             }

   \date{Received -; accepted -}

% \abstract{}{}{}{}{} 
% 5 {} token are mandatory
 
  \abstract
   {Time-dependent potentials are common in galactic systems that undergo significant evolution, interactions, or encounters with other galaxies, or when there are dynamic processes such as star formation and merging events. Recent studies show that an ensemble approach along with the so-called snapshot framework in the theory of dynamical systems provide a powerful tool to analyze the time-dependent dynamics. }
   {In this work, we aim to explore and quantify the phase space structure and dynamical complexity in time-dependent galactic potentials consisting of multiple components.}
   {We applied the classical method of Poincaré surface of sections to analyze the phase space structure in a chaotic Hamiltonian system subjected to parameter drift. This, however, makes sense only when the evolution of a large ensemble of initial conditions is followed. Numerical simulations explore the phase space structure of such ensembles while the system undergoes a continuous parameter change. The pair-wise average distance of ensemble members allowed us to define a generalized Lyapunov exponent, which might also be time-dependent, to describe the system stability.}
   {We provide a comprehensive dynamical analysis of the system under circumstances where linear mass transfer occurs between the disk and bulge components of the model.}
  {}


   \maketitle
%
%-------------------------------------------------------------------
\nolinenumbers
\section{Introduction}

The theory of stellar orbits in a galactic potential is a crucial aspect of galactic dynamics and astrophysics. It involves understanding how stars and other celestial objects move under the influence of combined gravitational forces within a galaxy \citep{Binney2008,Contopoulos2002}.
The motion of stars is influenced not only by the central mass concentration (such as a supermassive black hole in the case of many galaxies) but also by the combined gravitational influence of all the stars and dark matter. This leads to complex and chaotic orbital motions.
Numerically solving the equations of motion for stars in this potential allows to model and predict the overall behavior of the stellar population and interstellar medium from a dynamical standpoint.

Galaxies can be classified based on the symmetry of their mass distribution, usually as axisymmetric \citep{Toomre1963,Miyamoto1975} 

and non-axisymmetric potentials 

\citep{Binney1982,Contopoulos1990,Cincotta1996}. The shape of the potential significantly affects the stellar orbits.

The study of stellar orbits is closely related to the measurement of a galaxy's rotation curve, which plots the orbital velocity of stars as a function of distance from the galactic center. The rotation curve can provide valuable information about the mass distribution within the galaxy \citep{2001Verheijen,Gonzalez2010}.

%----------------------------------------------------------
In a galactic potential, the orbits of stars and other celestial objects can be broadly classified into two categories: regular orbits and chaotic orbits \citep{Lichtenberg1992,Contopoulos2002}. These classifications depend on the initial conditions of the orbits and the underlying dynamics governed by the gravitational forces within the galaxy. The study of regular and chaotic orbits is a fundamental aspect of galactic dynamics and provides valuable insights into the overall structure, formation, and evolution of galaxies. 
Understanding the stability of orbits is essential to explain the observed structures and dynamics of galaxies. \citep{Henon1964,Benettin1980,Skokos2010,Froeschle1997,Cincotta2000,Laskar1990,Gottwald2004}.

The understanding of the global structure of galaxies has rapidly improved with the development of many studies of rotation curves and mass density distributions calculated from observational data, which are relatively easy to interpret and decompose into different galaxy components. The very first models were built from observational data from our own Galaxy, the Milky Way \citep{Schmidt1956,Toomre1963,Kuzmin1956}, followed by various milestones in radio astronomy and more by the Hipparcos project \citep{Hipparcos}, several updated mass density models \citep{Bahcall1980,Caldwell1981,Miyamoto1975}, based on data from stars in and near our galaxy and other galaxies. These models are usually completed in the form of a potential, describing the gravitational interactions in the interior of galaxies in a compressed form, thus easily modeling the motions of different stars, clusters, and other stellar objects.
\par In recent years the re-evaluation and refinement of these models was possible due to more sophisticated and accurate observational tools, of which one of the most important and ambitious projects was the Gaia Survey \citep{GAIAORIGIN}. One of the Gaia Survey goals was to map out the stellar population and obtain valuable parameters, including kinematic data, in great accuracy in order to receive an even more realistic picture of our own galaxy \citep{GAIA2018}. 

Chaotic orbits arise when the gravitational potential is more complex, such as in regions with irregular mass distributions or significant gravitational perturbations from nearby massive objects.
Furthermore, chaotic orbits can arise due to orbital resonances induced by different gravitational influences and other kinematic characteristics of the orbits in a given system, which result in orbital periods related by simple integer ratios between the objects. In grand schemes, this may lead to the formation of structures in the phase space which give valuable information about the motion of an ensemble in certain regions of the system \citep{MALHOTRA1998}.

%----------------------------------------------------------

Motion in a time-dependent galactic potential refers to the study of how stars and other celestial objects move in a galaxy when the gravitational potential is not static but changes with time. Time-dependent potentials describe galactic systems that undergo significant evolution, interactions, or encounters with other galaxies, or when dynamic processes such as star formation and merging events take place inside certain regions.
Analyzing motion in a time-dependent galactic potential can be more challenging than in a static potential because the gravitational forces experienced by objects change over time (\cite{Caranicolas1996,Caranicolas2003,Manos2013,Manos2014,Zotos2011,Zotos2012,Zotos2014,Zotos2020}).  As a result, the orbits of stars can become more complex and less predictable on larger timescales. 

There are many studies in the recent past that have dealt with mass exchange between different parts of a galaxy. For example \citet{Caranicolas2003} and \citet{Zotos2012}
investigated stellar dynamics in cases where low angular momentum stars are scattered through the nucleus into the halo. This process transfers mass from the disk to the nucleus, while the total mass of the galaxy remains unchanged. Their model involves exponential mass transfer as a time-dependent parameter change in the potential. \citet{Manos2013} modeled the results of self-consistent N-body simulations by introducing an analytical time-dependent potential with linear mass transfer between the disk and the bar. They showed that an increase in the mass of the bar leads to more chaotic dynamics.

When a Hamiltonian system is subjected to a parameter shift, this means that some parameters in the Hamiltonian function are changed with time. These parameter shifts can occur due to various reasons, such as external influences or changing physical conditions in the system. The specific form of the parameter shift determines how the Hamiltonian depends on time and, in turn, affects the system's dynamics \citep{Lichtenberg1992}. Different parameter shifts can lead to various interesting phenomena in Hamiltonian systems such as an adiabatic and nonadiabatic parameter change, resonances, and chaos.

Substantial work was done in the past on galactic systems with potentials containing components undergoing perturbations. One such work was the analysis done by \cite{Pichardo2003...582..230P}, who looked at changes in systems where an axisymmetric galactic potential was perturbed by three-dimensional galactic spiral arms, which were added as a superposition of inhomogeneous oblate spheroids, introducing a periodic change into the system. This resulted in the emergence of nonlinear effects in the orbits of test particles which showed significant changes in the overall behavior of the system, which can mainly be seen in the Poincaré surface of sections shown in the work mentioned above. 

An active research area in chaos theory is the study of systems subject to monotonic nonadiabatic parameter changes. In this context, a comprehensive picture of these type of dissipative (i.e., frictional) system has emerged over the last three decades, motivated by climate change. These systems can be considered as simple models of the Earth's climate when a parameter (e.g. carbon dioxide concentration) varies monotonically. Here, it is shown that a long time study of a single trajectory is not representative; instead, the relevant phase space structure, the so-called snapshot chaotic attractor, is revealed by tracing an extended trajectory ensemble \citep{Yu1990,Sommerer1993,Chekroun2011,KT20}. In contrast, the theory of chaotic Hamiltonian (or conservative) systems subject to monotonic parameter changes has been a completely unknown field, but in the last years some discoveries have been made in this area, allowing for several methods to be developed to describe such systems \citep{Janosi2019,Janosi2021,Janosi2022}; for a review, readers can refer to \cite{Janosi2024}. One of the most important discoveries is the study of so-called snapshot tori, which are the relevant ensembles in the phase space. These are generalizations of the so-called Kolmogorov-Arnold-Moser (KAM) tori of stationary Hamiltonian systems, and can be defined as closed curves in the phase space of systems with varying parameters, but their shape is not constant, and over time their dynamics can become chaotic through a breakup process. Another important aspect is the generalization of Lyapunov exponents, through the calculation of the so-called ensemble-averaged pairwise distance (EAPD).

In this work, we aim to explore and quantify the phase space structure and dynamical instability in time-dependent galactic potentials simulating mass exchange between the disk and the central bulge. We also aim to demonstrate the efficiency of the recently proposed ensemble description and snapshot method 
in time-varying Hamiltonian dynamics. 

The paper is organized as follows. Section~\ref{sec:2} presents the model we used and the parameters obtained from radial velocity fits. Section \ref{sec:3} is devoted to the phase space analysis of the steady-state and time-dependent stellar dynamics. In Section \ref{sec:4} we summarize our results and draw conclusions on the implications.

%==================================================================

%--------------------------------------------------------------------
\section{The model}\label{sec:2}

\subsection{Potential profiles for the components}
In this work the analysis was done on Milky Way-like galactic potential models. These models can be described as a combination of three separate potentials, each representing a prominent component of a galactic system: a galactic disk, a central bulge and a dark matter halo component.
\subsubsection{Galactic disk}
For several decades the most well-known and commonly used galactic potential model was the one described by \cite{Miyamoto1975}, which is a mathematically simple, yet good approximation when it comes to the numerical analysis of such systems.
For the disk components for our constructed models an extended Miyamoto-Nagai potential was chosen, which is derived from the Poisson equation \citep{Vogt2005}, and was used in the analysis of three-dimensional axisymmetric potential models for the Milky Way by \cite{Barros2016}. \par 
The equation for the extended Miyamoto-Nagai disk potential is as follows:

\begin{equation}\label{gen_miyamoto}
    \begin{split}
    \Phi_{d}(R,z)=&-\dfrac{A}{\sqrt{R^2+(a+\bar{z})^2}}\bigg (1+\dfrac{a(a+\bar{z})}{R^2+(a+\bar{z})^2}-\\
    &-\dfrac{1}{3}\dfrac{a^2\big[R^2-2(a+\bar{z})^2\big]}{\big[R^2+(a+\bar{z})^2\big]^2} \bigg),
    \end{split}
\end{equation}
where $R$ and $z$ are the cylindrical coordinates, $A$ is the mass of the disk component multiplied by the gravitational constant $G$, while $\bar{z}=\sqrt{z^2+b^2}$. The parameter $a$ is the radial scaling length, while $b$ is the scaling height, as it is commonly denoted in the Miyamoto-Nagai model.  
\subsubsection{Bulge}
For the central bulge, which is commonly approximated as a spheroidal component, we adopted the Hernquist potential as a simple alternative \citep{Hernquist1990}. 
\par 
The potential takes the following form:
\begin{eqnarray}
\begin{gathered}\label{hernquist}
    \Phi_{b}(R,z)=-\dfrac{C}{\sqrt{R^2+z^2}+c},
\end{gathered}
\end{eqnarray}
where $C$ again is a mass multiplied by the gravitational constant $G$, while $c$ is the scaling radius of the spheroidal bulge.
\subsubsection{Dark matter halo}
For the dark matter halo component a Navarro-Frenk-White (NFW) profile \citep{NFW1996} was chosen, written in terms of the scaling radius $r_s$ and the enclosed mass $M_{s}$ \citep{Sanderson_2017}, instead of the commonly used form determined by the scale density $\rho_s$ or virial radius $r_{vir}$ and mass $M_{vir}$, in order to keep the potential components mathematically consistent.
The equation for the scaling mass is as follows:
\begin{eqnarray}
\begin{gathered}\label{mass}
    M_s=4\pi \rho_s r_s^3(\ln 2-1/2),
\end{gathered}
\end{eqnarray}
and the NFW profile in terms of $M_s$ can be written in the following form using the cylindrical coordinate system:
\begin{eqnarray}
\begin{gathered}\label{nfw}
    \Phi_{DM}(R,z)=-\dfrac{D}{\ln{2}-1/2}\dfrac{\ln(1+\sqrt{R^2+z^2}/d)}{\sqrt{R^2+z^2}}, 
\end{gathered}
\end{eqnarray}
where $D = GM_s$ is the enclosed mass multiplied by the gravitational constant $G$ and $d=r_s$. \par 
\vspace{0.2cm}
The overall model of a Milky Way-like disk galaxy is simply the sum of the disk, bulge, and dark matter halo potentials written in Eqs.~(\ref{gen_miyamoto},~\ref{hernquist},~\ref{nfw}):
\begin{eqnarray}
\begin{gathered}\label{allpot}
    \Phi(R,z) = \Phi_d+\Phi_b+\Phi_{DM}. 
\end{gathered}
\end{eqnarray}

\subsection{Equations of motion}

The motion of a test particle (star or interstellar object) can be described by the following Lagrangian, with $L_z = p_\phi$ being a constant representing the angular momentum component around the $z$-axis since the models under consideration exhibit axial symmetry (more precisely, cylindrical symmetry):

\begin{eqnarray}
\begin{gathered}
\mathcal{L} = \frac{1}{2}\bigg(\dot{R}^2 + \dot{z}^2 \bigg) - \Phi(R, z),
\end{gathered}
\end{eqnarray}
where $(R, z)$ represent the cylindrical coordinates, and we incorporate the summed potential given by Eq.~(\ref{allpot}) into the expression for $\Phi(R, z)$. The corresponding generalized momenta are
\begin{eqnarray}
\begin{gathered}
p_R = \dot{R}, \quad p_z = \dot{z},
\end{gathered}
\end{eqnarray}
and thus, the Hamiltonian can be written as
\begin{eqnarray}\label{hamilton}
\begin{gathered}
\mathcal{H} = \frac{1}{2}(p_R^2 + p_z^2 ) + \Phi_{\text{eff}}(R, z),
\end{gathered}
\end{eqnarray}
where the effective potential $\Phi_{\text{eff}}$ is given by
\begin{eqnarray}
\begin{gathered}
\Phi_{\text{eff}}(R, z) = \frac{L_z^2}{2R^2} + \Phi(R, z),
\end{gathered}
\end{eqnarray}
where $L_z$ is the $z$-component of the angular momentum.

Thus, out of the original six generalized coordinates, only four remain, allowing us to define 4 first-order differential equations describing the motion:

\begin{eqnarray}\label{eqm}
\begin{gathered}
\dot{R} = p_R, \
\dot{z} = p_z, \\
\dot{p_R} = \frac{L_z^2}{R^3} - \frac{\partial \Phi(R, z)}{\partial R}, \
\dot{p_z} = -\frac{\partial \Phi(R, z)}{\partial z}.
\end{gathered}
\end{eqnarray}

\subsection{Fit of radial velocity profile}\label{rotation curve}
%-------------------------------------- Two column figure (place early!)

%

The rotation curve equation can be obtained by the following general relation:
\begin{eqnarray}
\begin{gathered}\label{gen_miyamoto_v}
    v_c(R)=\bigg(R\dfrac{\partial \Phi(R,z)}{\partial R}\bigg\vert_{z=0}\bigg)^{\frac{1}{2}}.
\end{gathered}
\end{eqnarray}

This equation is needed for determining the parameters $a,b,c,d, A, C$ and $D$. Two main sources of data were used as a combined data set: one is the grand rotation curve of the Milky Way compiled by \cite{Sofue2012, Sofue_2013}, while the other one is the collection of rotational velocity data obtained using Classical Cepheids by \cite{Mroz2019}, ensuring that the inner parts of the rotation curve around the range of $R=0-20 \; \mathrm{kpc}$ is well-captured. After that the parameter space can be explored to find the best-fitting parameters for constructing a model representing a Milky Way-like disk galaxy. 
 
\begin{figure}[!htb]
\centering
\includegraphics[width=\linewidth]{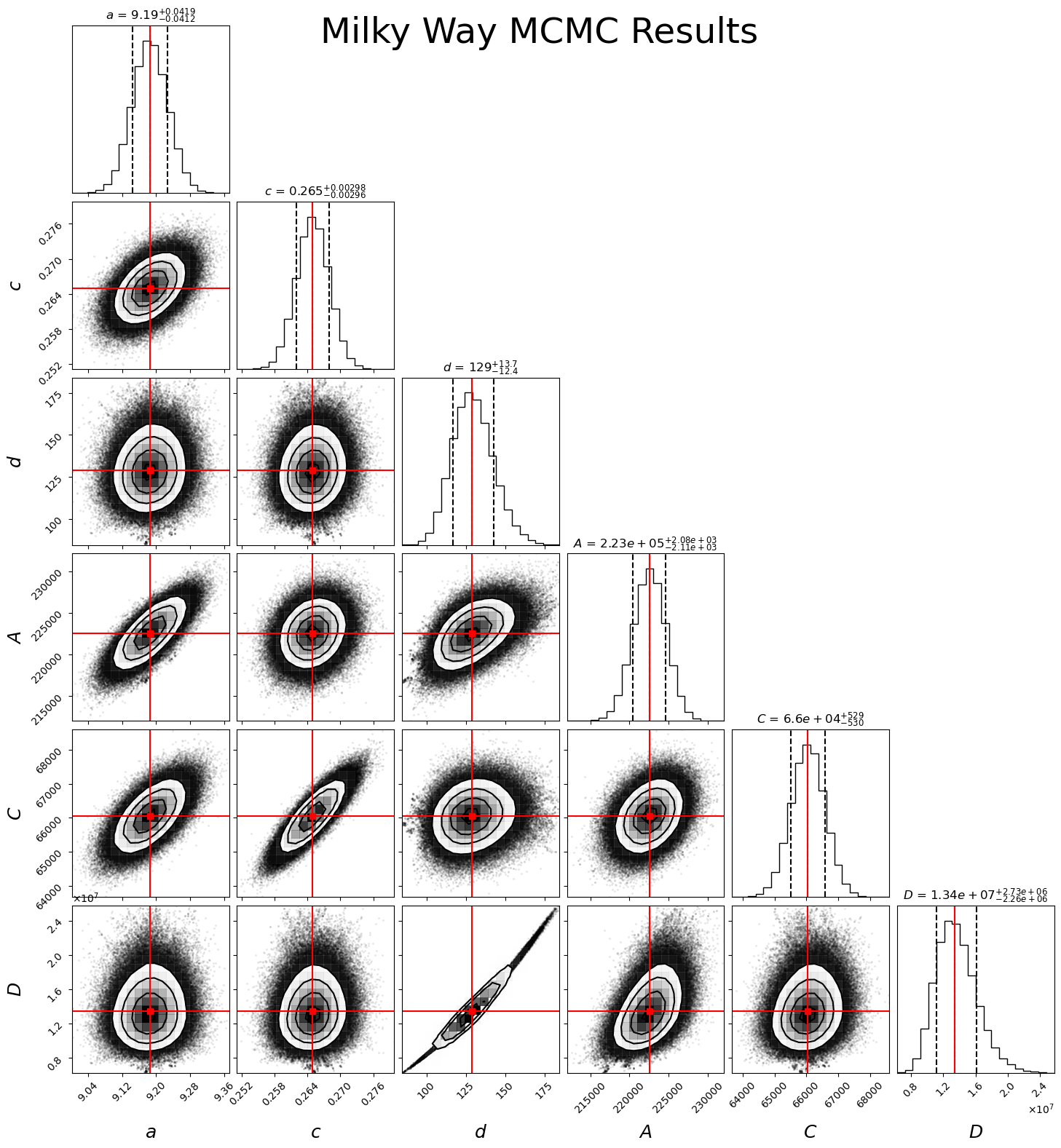}
\caption{MCMC results on the gathered rotational velocity data illustrated in a corner plot. As expected, the highest correlations can be seen between the scaling distances and mass components corresponding to the same potential components. The red dots in the two-dimensional projection, along with the red lines in the one-dimensional projection, show the Maximum Likelihood estimates for the parameters, along with the black lines showing the median and $1\sigma$ confidence interval results.}
\label{fig:mcmc}%
\end{figure}

The procedure to find the best parameters was done in two steps:
 \begin{itemize}
     \item First, the rotation curve Eq.~(\ref{gen_miyamoto_v}) was fitted using Truncated Newtonian Method, or TNC, \citep{Dembo1983TruncatednewtonoAF} for the parameters $a,c,d,A,C,D$ to receive approximate values in preparation for the next step. The initial parameter guesses of the scaling distances and masses, along with the lower and upper limits for the parameter space, for the TNC fit were obtained from the results of the DR3 + data analysis of \cite{Labini_2023} and the works of \cite{Paul2011} and \cite{McMillan_2016}. The fits were done using the \textit{lmfit} Python library \citep{Newville2015_11813}.
     
     \item After that, the fitted parameters from the TNC were used as initial conditions for an affine invariant Markov chain Monte Carlo (MCMC) model \citep{Goodman2010...5...65G}. The change in disk scale height has been shown to give minimal contribution to the main features of such models, as was shown in \cite{Paul2011}, so it was chosen as a constant $b=0.25 \; \mathrm{kpc}$, being the lowest used scale height in the aforementioned work. The software of choice used for performing the MCMC samplings was \textit{emcee} \citep{Foreman_Mackey_2013}. The number of walkers was chosen to be 1000, with a sampling time of $t_{smp}=10^6$ and a burning time of 5 times the time of highest autocorrelation time $t_{ac} \approx 200$. The same lower and upper constraints were implemented lmfitfor the parameter space in this analysis as in the previous step. The results of this analysis can be seen in Fig.~\ref{fig:mcmc}, while the obtained parameters are summarized in Table~\ref{params_table}.
 \end{itemize}
 
\begin{table*}[!htb]
\centering
\caption{Parameters obtained from the TNC and MCMC methods.}
\begin{tabular}{c|
>{\columncolor[HTML]{FFFFFF}}c |c|
>{\columncolor[HTML]{FFFFFF}}c 
>{\columncolor[HTML]{32CB00}}c |}
\cline{2-5}
\cellcolor[HTML]{FFFFFF}                                                & \cellcolor[HTML]{9B9B9B}                                & \cellcolor[HTML]{9B9B9B}                             & \multicolumn{2}{c|}{\cellcolor[HTML]{9B9B9B}MCMC}                                                                   \\ \cline{4-5} 
                                                                        & \multirow{-2}{*}{\cellcolor[HTML]{9B9B9B}Initial} & \multirow{-2}{*}{\cellcolor[HTML]{9B9B9B}TNC}        & \multicolumn{1}{c|}{\cellcolor[HTML]{C0C0C0}Median}                                   & \cellcolor[HTML]{C0C0C0}MLE \\ \hline
\multicolumn{1}{|c|}{\cellcolor[HTML]{9B9B9B}$a$ {[}kpc{]}}             & 4.5                                                     & \cellcolor[HTML]{FFFFFF}9.150                        & \multicolumn{1}{c|}{\cellcolor[HTML]{FFFFFF}$9.190^{+0.042}_{-0.041}$}              & 9.179                       \\ \hline
\multicolumn{1}{|c|}{\cellcolor[HTML]{9B9B9B}$c$ {[}kpc{]}}             & 0.25                                                    & \cellcolor[HTML]{FFFFFF}0.253                        & \multicolumn{1}{c|}{\cellcolor[HTML]{FFFFFF}$0.265^{+0.003}_{-0.003}$}            & 0.265                       \\ \hline
\multicolumn{1}{|c|}{\cellcolor[HTML]{9B9B9B}$d$ {[}kpc{]}}             & 183                                                     & \cellcolor[HTML]{FFFFFF}124.9                          & \multicolumn{1}{c|}{\cellcolor[HTML]{FFFFFF}$129.0^{+13.7}_{-12.4}$}                    & 126.3                        \\ \hline
\multicolumn{1}{|c|}{\cellcolor[HTML]{9B9B9B}$A$ {[}kpc km s$^{-1}${]}} & $1.29\cdot10^{5} $                                     & \cellcolor[HTML]{FFFFFF}$2.29 \cdot 10^{5}$          & \multicolumn{1}{c|}{\cellcolor[HTML]{FFFFFF}$(2.23^{+0.021}_{-0.021})\cdot 10^{5}$} & $2.22\cdot 10^{5}$          \\ \hline
\multicolumn{1}{|c|}{\cellcolor[HTML]{9B9B9B}$C$ {[}kpc km s$^{-1}${]}} & $8.6\cdot10^{4}$                                     & \cellcolor[HTML]{FFFFFF}$6.48 \cdot 10^{4}$          & \multicolumn{1}{c|}{\cellcolor[HTML]{FFFFFF}$(6.60^{+0.053}_{-0.053})\cdot 10^{4}$} & $6.60\cdot 10^{4}$          \\ \hline
\multicolumn{1}{|c|}{\cellcolor[HTML]{9B9B9B}$D$ {[}kpc km s$^{-1}${]}} & $6.9\cdot 10^{6} $                                    & \cellcolor[HTML]{FFFFFF}$1.23 \cdot 10^{7}$          & \multicolumn{1}{c|}{\cellcolor[HTML]{FFFFFF}$(1.34^{+0.27}_{-0.23})\cdot 10^{7}$}     & $1.29\cdot 10^{7}$          \\ \hline
\end{tabular}
\tablefoot{The final parameters used in the fitting of the rotation curve are indicated in green. The scaling height $b$ was kept at a constant $0.25 \; \mathrm{kpc}$.}
\label{params_table}
\end{table*}

\begin{figure}[!htb]
    \centering
    \includegraphics[width=\linewidth]{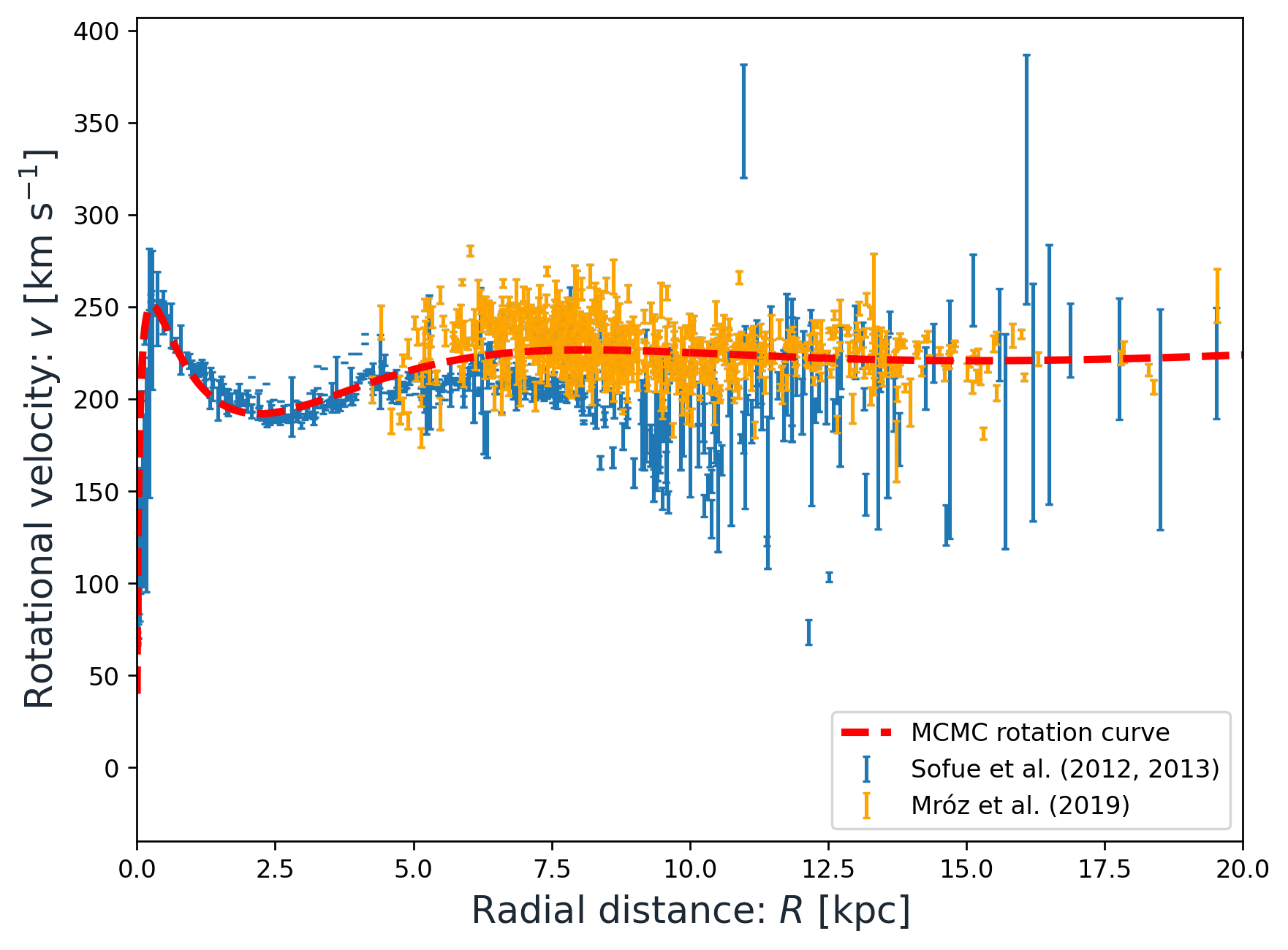}
    \caption{Rotation curve fitted to the observational data indicated in the legend, using parameters obtained from the MCMC analysis of Fig.~\ref{fig:mcmc} and shown in Table \ref{params_table}.}
    \label{fig:rotation_curve}
\end{figure}

The rotation curve resulting from the fit is shown in Fig.~\ref{fig:rotation_curve}. The fitting was done on the joint datasets of \cite{Sofue2012, Sofue_2013} (blue) and \cite{Mroz2019} (orange), and shows quite good agreement with the observations. Although some of the resulting parameters deviate from the most recent estimates for the Milky Way \citep{10.1093/mnras/stae034, Labini_2023, 2019A&A...621A..56P}, they remain within acceptable bounds for constructing a Milky Way-like galactic potential model. These parameters are adequate for the upcoming analysis, where the primary goal is to have a model which may reproduce the rotation curve in the inner regions of the galactic model as seen on Fig.~\ref{fig:rotation_curve}.

\section{Results}
\label{sec:3}

\subsection{Time-independent case}\label{sub:stationary}
The energy of a trajectory initiated from a particular set of initial conditions can be determined by evaluating the Hamiltonian function (\ref{hamilton}) using the given initial conditions. Assuming a drift-free case, the energy of the trajectory is conserved at all times. Therefore, we can express it as follows:

\begin{eqnarray} \label{Ham}
\begin{gathered}
\mathcal{H}(R, z, p_R, p_z) = \mathcal{H}(R_0, z_0, {p_R}_0, {p_z}_0) = H.
\end{gathered}
\end{eqnarray}

Merely tracing trajectories is far from sufficient for an extensive examination of a system, where a deeper understanding of a specific galaxy model's morphology and the structures within it can be obtained. Investigating trajectories of a system with multiple initial conditions using physical and phase space plots, depending on the number of initial conditions, may not be efficient or even feasible. Therefore, it is required to continue the investigations using the Poincaré SOSs method \citep{ott93_2, Teschl_ordinary}. 

Each intersection is registered under the conditions $z=0$ and $p_z>0$.
To clarify, from this point on all the units for momenta, angular momenta, Hamiltonian energy and parameters $A, B, C$ will be indicated per unit mass, making them specific quantities.

In order to obtain the initial conditions for the simulations we first need to find the circular orbit at $R_{circ}=8.249 \; \mathrm{kpc}$ and $z_{circ}=0.0208\; \mathrm{kpc}$, which is the estimated location of our Sun in the galactocentric coordinate system from \cite{Gravity2021}, and calculate the radial velocity $V_R \approx 226.77 \; \mathrm{km \, s^{-1}}$ from the derived rotation curve. This value can then be used to acquire the $z$-component of the angular momentum (per unit mass) $L_z=V_RR$. After that, the reference Hamiltonian energy (per unit mass) for a circular orbit $H_0 = -5.26 \cdot 10^5 \; \mathrm{km^2 \; s^{-2}}$ can be calculated using Eq.~(\ref{hamilton}), with ${p_R}_0,{p_z}_0=0$.

By lowering the $L_z$ parameter it is possible to create an ensemble of orbits which have their Hamiltonian energies distributed to the $p_R, p_z$ velocity components, resulting in orbits which are misaligned with the galactic plane. Stellar streams can have such orbits, and in the last couple of years they have been proven to be useful tools for measuring the Milky Way's gravitational potential \citep{Sanderson_2017, walder2024probing}. Figure \ref{fig:stac2} of the Appendix illustrates the effect of different $L_z$s while keeping $H_0$ fixed.

It is possible to find initial parameter sets by setting $p_{R_0}=0$ and calculating the maximum permitted value for the velocity in the $z$-direction $p_{z_{max}}$ using Eq.~(\ref{hamilton}), with known $H$ and a chosen $L_z$, and scanning the phase space with an arbitrary resolution in the range of $p_{z_0}=[0; p_{z_{max}}]$.
By simultaneously going through the range of radial distances $R_0=[0;R_{circ}]$ it is possible to find all the possible orbits in the phase space with a given $H_0$ and $L_z$ parameter values. 

A total of 400 initial conditions were launched in each chosen $L_z, H$ values. The numerical integrations were performed using the NumbaLSODA integrator \citep{numbalsoda}, which uses the LSODA algorithm with automatically switching between stiff and nonstiff methods during the integration depending on which of the two can more likely solve the equation in most efficient manner, which can especially be beneficial in cases where the nature of the problem is unknown or may change drastically during the chosen time intervals \citep{LSODAOrigin}.
The simulations in the time-independent case were run for $t=5000 \; \mathrm{Gyr}$ with a resolution of $\text{d}t=10^{-4}\; \mathrm{Gyr}$, which is sufficient to determine the intersections within a certain time interval, to obtain detailed Poincaré SoSs .

\begin{figure}[!htb]
    \centering
    \includegraphics[width=\linewidth]{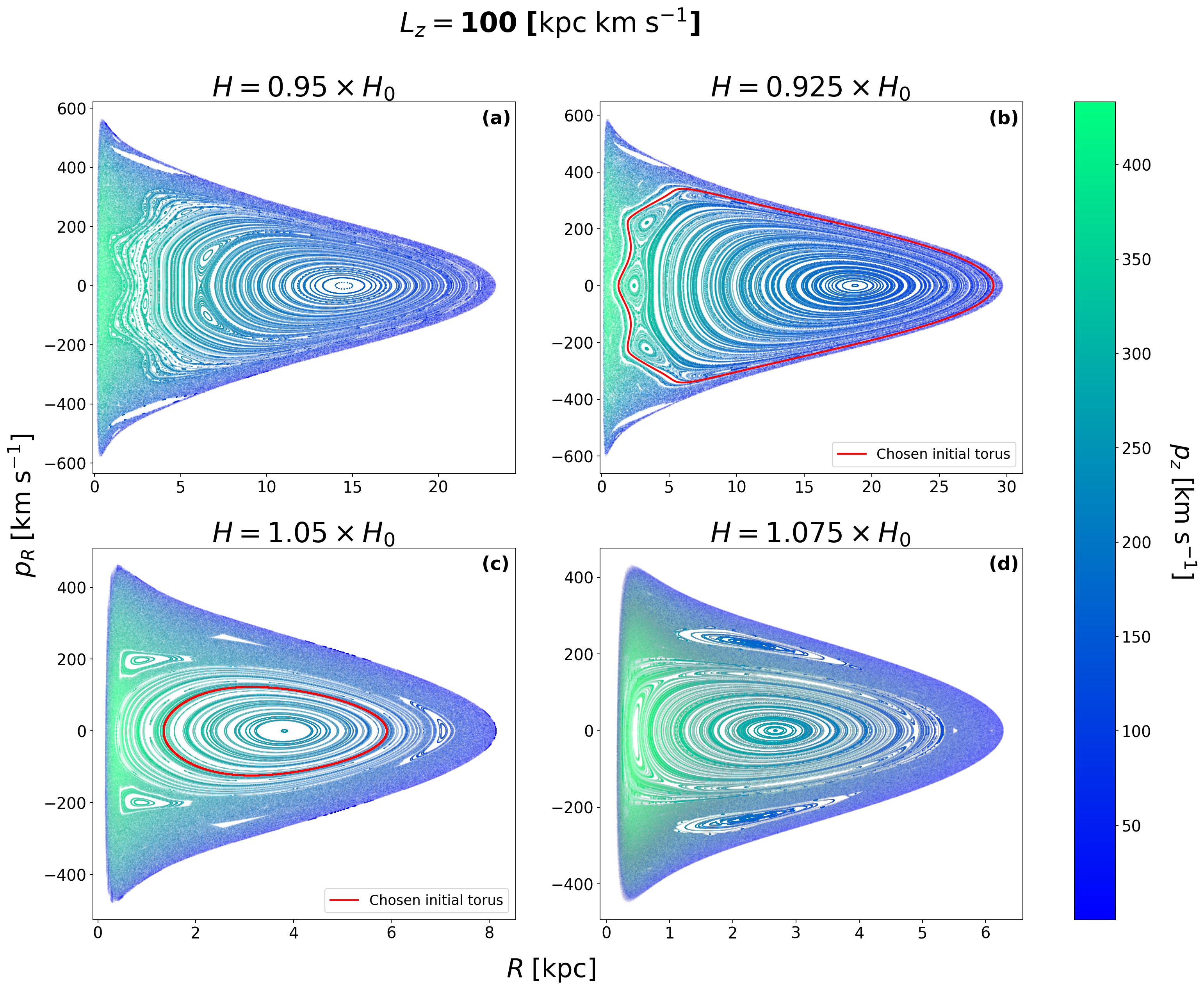}
    \caption{Phase portraits of Eq.~\eqref{eqm} for different values of $H$ with constant $L_z=100$. The trajectories are plotted on the $(R,p_R)$ plane, following the Poincaré cut at $z = 0$, and the value of $p_z$ obtained from \eqref{Ham}, indicated at each point by the colormap. The red KAM tori belong to initial conditions $R_0 =1.34, \; {p_R}_0 =  0$ (panel b) and $R_0 = 2.96,\; {p_R}_0 = 267.37$ (panel c).} 
    \label{fig:stac}
\end{figure}

Figure \ref{fig:stac} show the phase space plots in the $(R,p_R)$ plane, with a colormap representing the value of $p_z$ at each point, obtained from Eq.~\eqref{Ham} with $z = 0$.
With a chosen reference of $L_z=100$ we illustrate the phase portrait for four different values of constant $H$. The red curves are highlighted KAM tori of initial conditions $R_0 =1.34, \; {p_R}_0 = 0$ (Fig.~{\ref{fig:stac}}b) and $R_0 = 2.96 , \; {p_R}_0 = 267.37$ (Fig.~{\ref{fig:stac}}c). 
The analysis of the time-dependent dynamics to follow in the next section will be done on these curves.

One can observe that the system possesses a typical divided phase space with the coexistence of four types of dynamics. Quasi-periodic motion occurs on the closed curves, that is, the KAM tori, where trajectories densely fill these shapes. These tori form concentric islands, where the radius of the shapes depends on the potential energy of the trajectories, which is conserved within a curve. In the center of such islands one finds so-called elliptic fixed points, representing periodic trajectories.

The third type of dynamics is chaos, which manifests in this illustration as dense bands of irregularly mixed trajectories, called the chaotic sea, observable along the edge of the phase space, between the large central and several smaller islands. Motion within this region is ergodic, meaning that trajectories uniformly cover the chaotic sea, and have equal probability to reach any parts of it. Here the dynamics is unpredictable, and trajectories, even if they were very close initially, tend to diverge at an exponential rate.

Although the shown phase portraits do not capture this explicitly, it is known from chaos theory (see e.g. \cite{ott93_2}) that a peculiar phenomenon occurs on the border between the island of KAM tori and the chaotic sea. In a narrow band outside the outermost KAM torus the trajectories do not form any comprehensive pattern, they are scattered just like in the chaotic sea, however their dynamics is not yet chaotic, they do not diverge exponentially. Orbits are able to detach from this region into the chaotic sea (and vice versa), however the transport of trajectories in this direction was found to follow a power-law \citep{channon1980,karney1983,meiss1983,Cristadoro2008,Altmann2013}, which is much slower than the ergodic wandering in the chaotic sea. Because of this, chaotic trajectories can be "trapped" inside these regions and "stick" to the outermost torus for long times. This is the fourth type of motion present in the phase space, and is often referred to as sticky dynamics.

\subsection{Parameter drift and EAPD curves}\label{sec:EAPD}
The simulations conducted in Section \ref{sub:stationary} were calculated by keeping the parameters received from the model fittings fixed, assuming that the physical properties associated with these parameters, as well as those derived from them, do not change over time. The changes occurring in these systems are generally considered relatively slow, making these assumptions reasonable approximations for extrapolating behaviors of certain galaxies, which can be considered isolated and stable for a longer period of time when it comes to the stage of stellar evolution.

In this section, 
we assign a time dependence to parameters $A$ and $C$ by introducing linear drifts as 
\begin{equation}
    A(t) = A(0) - \mu t,     \;\;\;   C(t) = C(0) + \mu t,
    \label{drift}
\end{equation}
that is, the decrease in $A(t)$ and the increase in $C(t)$ has the {\em same} $\mu$ rate, that is, we simulate a {\em mass transfer} from the disk to the bulge, so that the overall mass of the galaxy is conserved.
The initial values $A(0), C(0)$ are read from Table \ref{params_table}, and the rate is chosen to be $\mu = \frac{3}{10}A(0) \cdot \frac{1}{T_{f}}$, with the final intersection time $T_f=200 \; \mathrm{Gyr}$.

Following the methods used in \cite{Janosi2019,Janosi2021,JKT21,Janosi2022,Janosi2024}, we have to accept that simulating individual trajectories is no longer reliable. Instead, one has to follow an {\em ensemble} of trajectories in order to obtain sufficient statistics. Furthermore, in Hamiltonian systems the shape of this ensemble is not arbitrary, they have to be initiated, for example, on KAM tori of the stationary system. In our case, we will focus on the two red tori highlighted in Fig.~\ref{fig:stac}.

The analysis includes studying the temporal evolution of the tori, whose shapes become time-dependent, and are called snapshot tori. It was found that at some point the dynamics of these snapshot tori becomes chaotic, indicated by a strong stretching in their shape. This is called the breakup of snapshot tori, and it ends with them becoming entrained into what is called the snapshot chaotic sea, the time-dependent image of the stationary chaotic sea, whose shape is also changing in time.

It should also be noted that the snapshot picture taken on consecutive Poincaré sections implies intersections at various time instants, which are in different stages of the parameter drift. Consequently, the energy does not remain constant. Therefore, what we depict is a projection of the $(R,p_R,p_z)$ phase space onto the $(R,p_R)$ plane, which has evolved from a stationary Poincaré section.

\par
\begin{figure}[!htb]
    \centering
    \includegraphics[width=\linewidth]{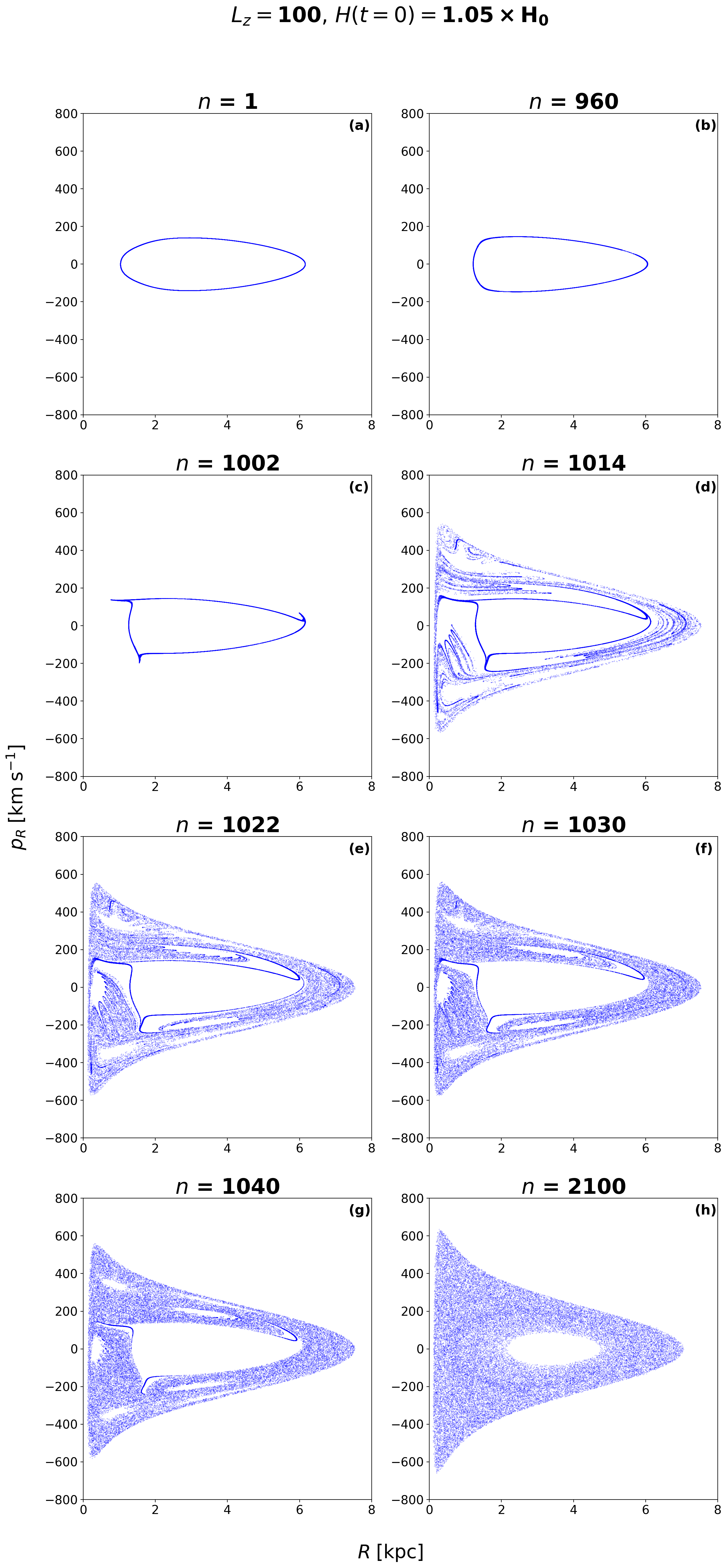}
    \caption{Time evolution of the red torus on Fig.~\ref{fig:stac}c. For around 1000 iterations it only deforms slightly, but at $n = 1002$ it starts to breakup, resulting in the dynamics of more and more trajectories of the snapshot becoming chaotic.
    }
    \label{fig:timdep_simple}
\end{figure}

\begin{figure}[!htb]
    \centering
    \includegraphics[width=\linewidth]{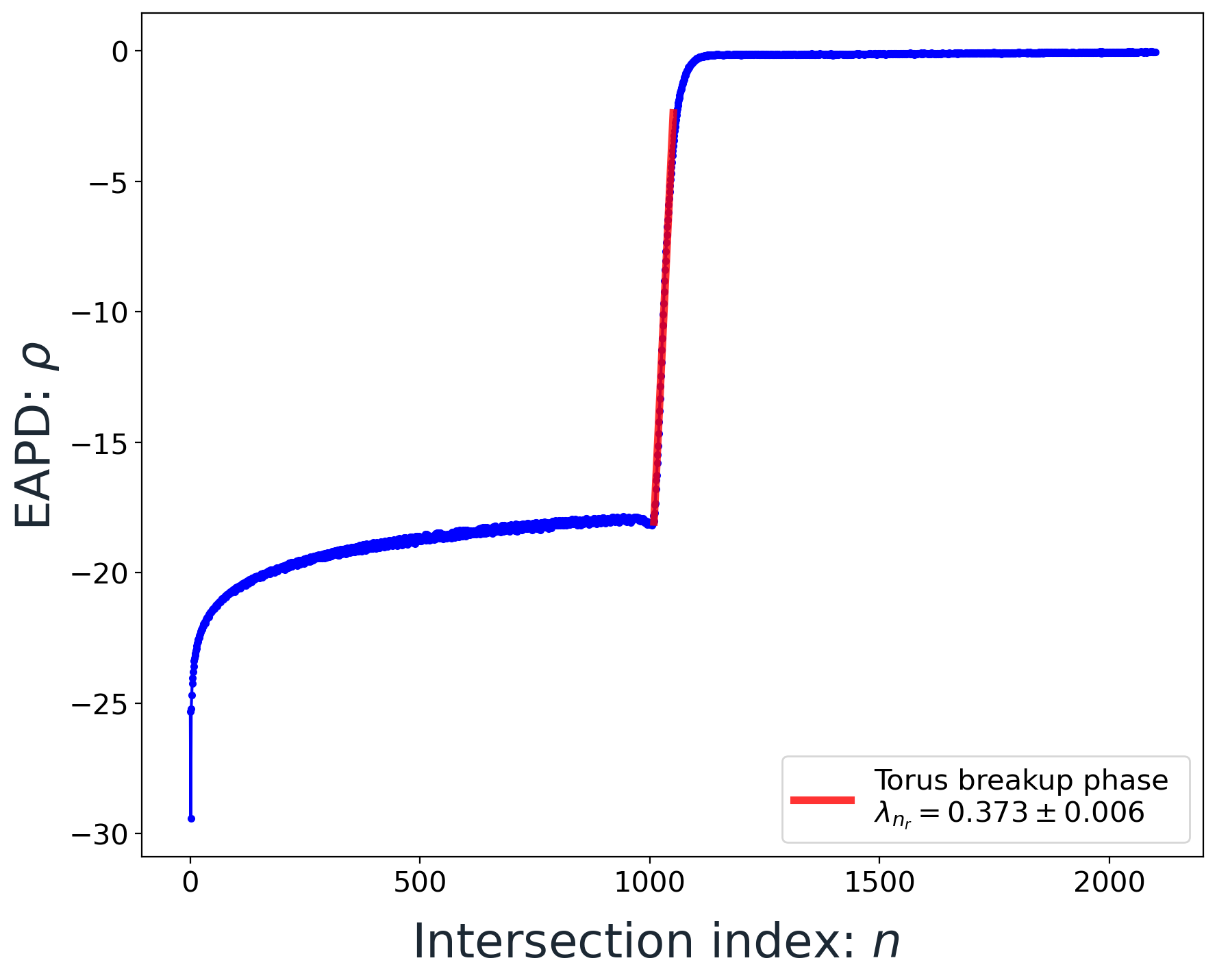}
    \caption{EAPD curve of Eq.~\eqref{EAPD}, evaluated for the snapshot torus of Fig.~\ref{fig:timdep_simple}. The Lyapunov exponent obtained from the fitting is $\lambda_{n_r} = 0.373 \pm 0.006$.}
    \label{fig:eapd_simple}
\end{figure}

\begin{figure}[!htb]
    \centering
    \includegraphics[width=\linewidth]{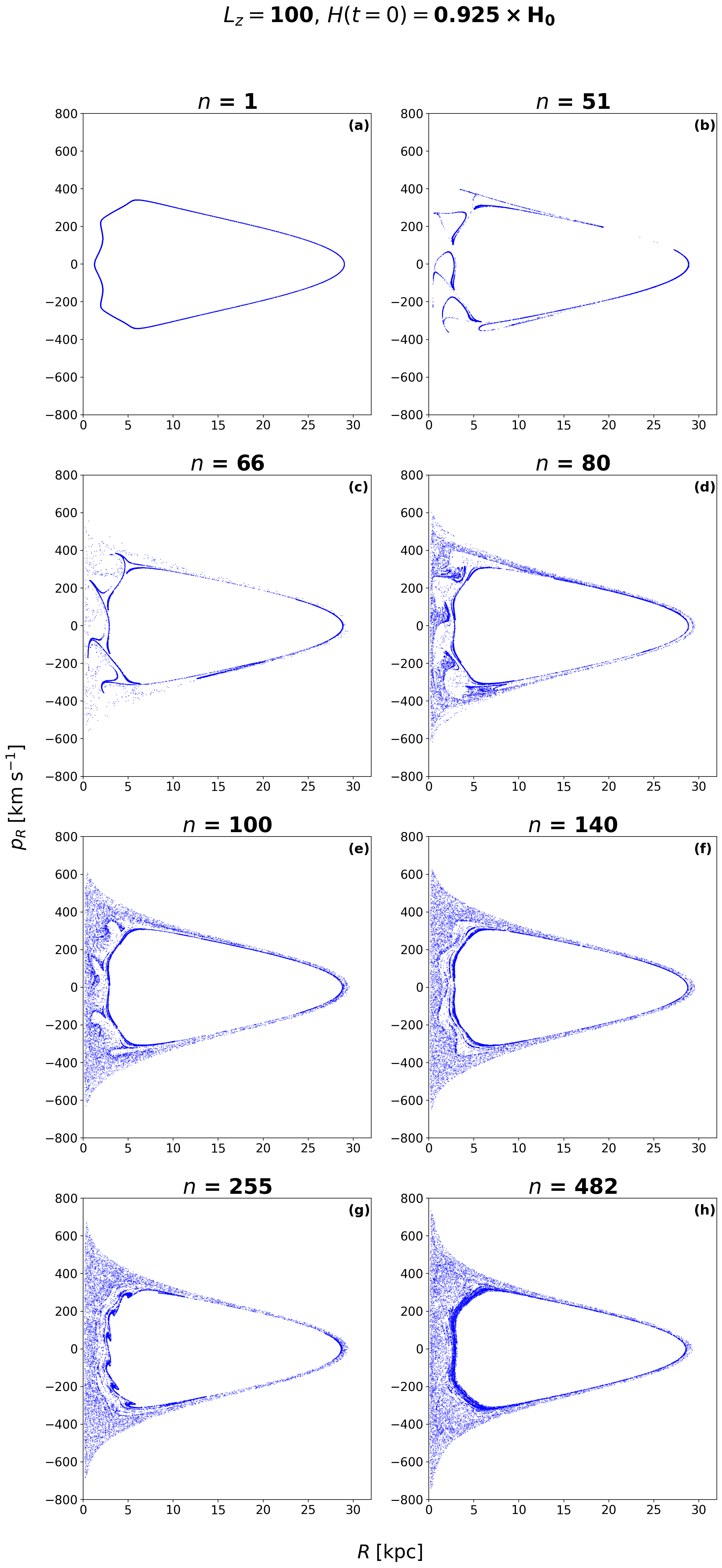}
    \caption{Time evolution of the red torus on Fig.~\ref{fig:stac}b. At $n = 50$ its points start to wrap around the snapshot tori initially belonging to the islands around the initial torus, until these breakup starting at $n = 100$. Points of the large snapshot torus are dispersed by instant $n = 482$.}
    \label{fig:timdep_intermediate}
\end{figure}

\begin{figure}[!htb]
    \centering
    \includegraphics[width=\linewidth]{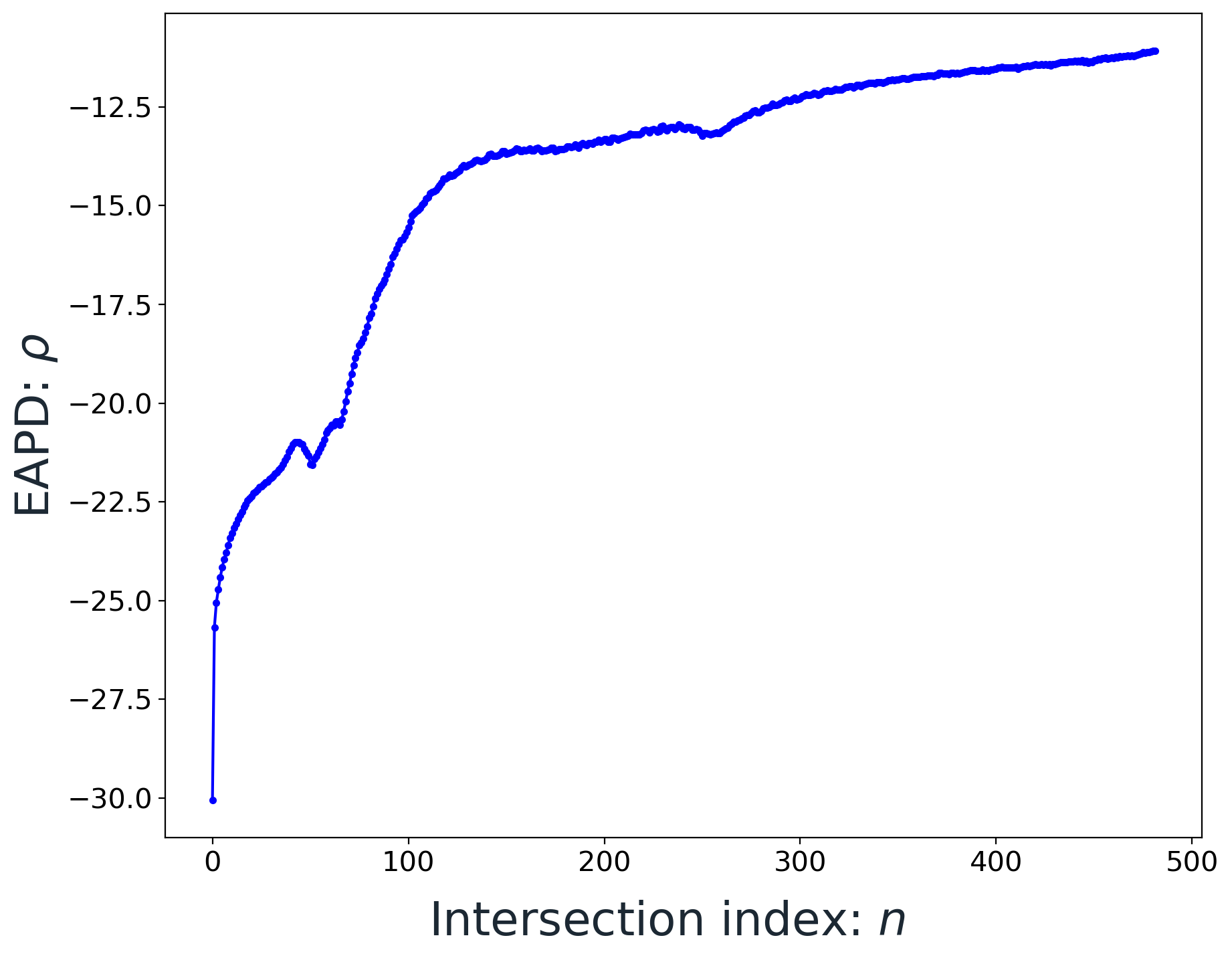}
    \caption{EAPD curve of Eq.~\eqref{EAPD} evaluated for the snapshot torus of Fig.~\ref{fig:timdep_intermediate}.
    }
    \label{fig:eapd_intermediate}
\end{figure}

 The most straightforward approach to elucidating our simulation results is through visualization. In Figs.~\ref{fig:timdep_simple}-\ref{fig:eapd_intermediate}, the time-dependent simulations showcase the time evolution of snapshot tori in the galactic model previously derived and discussed.  From such images we can easily acquire geometrical information about the general behavior of the evolution of the selected ensemble throughout the integration time.

In Fig.~\ref{fig:timdep_simple} we follow the snapshot torus whose initial image is the red KAM torus shown in Fig.~\ref{fig:stac}c. Its evolution is similar to those shown in \cite{Janosi2019,Janosi2021,Janosi2022,Janosi2024}: for some time, it does not appear to show any peculiar change in its dynamics, which can persist for longer times, as we see here at $n = 960$, although one can observe slight deformations in its shape, particularly on the left. However, this suddenly changes when the torus stretches strongly in multiple directions ($n = 1002$), followed by disintegration and, in the end, fully chaotic dynamics ($n = 2210$). 

For a quantitative analysis, we turn to the \textit{ensemble-averaged pairwise distance} (EAPD) method \citep{Janosi2019,Janosi2021,Janosi2022,Janosi2024}, which allows measuring the dynamical instability of time-dependent chaos. This method can be seen as a generalized form of the concept of Lyapunov exponents, where the fundamental behavior of the system (or significant parts of it) can be deduced from the ensemble average of distances between pairs of points in the phase space. In most analysis methods, these distances are evaluated on projections of the phase space.
The definition of EAPD is as follows:
\begin{eqnarray}\label{EAPD}
\begin{gathered}   
    \rho_n=\langle \ln{r_n} \rangle,
\end{gathered}
\end{eqnarray}
where the bracket denotes the average over the ensemble, $\ln{r_n}$ represents the natural logarithm at the $n$-th Poincaré section of the phase space distance $r_n$ between one point in the snapshot torus, and its assigned pair of initial distance $r_0 = \sqrt{2} \cdot 10^{-12}$, which is not on the snapshot torus itself but very close to it, which turns out to be a good approximation for our purposes. 
If the temporal change of the distances follows an exponential trend, we can fit a linear function defined on a natural semi-logarithmic scale, where the slope of the function is the so-called instantaneous Lyapunov exponent denoted by $\lambda$. 
In the above case $\lambda$ is constant, while in general, the instantaneous Lyapunov exponent is given as the derivative of the EAPD function,
\begin{equation}
    \lambda_n = \Dot{\rho}_n,
\end{equation}
where the dot denotes the time derivative evaluated in discrete time.
This approach 
%can be considered as the generalization of the  Lyapunov-exponent and 
allows us to make assumptions about the system in a manner similar to the traditional counterpart. If the $r_n$ distances diverge linearly during the time evolution, it indicates regular or periodic behavior, whereas in chaotic cases it results in exponential distance growth.

During the calculation of distances in phase space at the moments of intersection, it is important to take into account differences in units, since it is evident that the velocity components $p_R, p_z$ would overshadow the distances in $R$. In order to remedy this problem all the phase space dimensions are standardized using z-score transform, so that the standard deviations and means of phase space coordinates are comparable to each other. This also means that the distance growths in the following EAPD curves are not direct indications of the distances on the Poincaré sections, rather it is a normalized approach which, however, does capture the overall form of the evolution of distances in the actual phase space.

The EAPD function evaluated for the snapshot torus of Fig.~\ref{fig:timdep_simple} is shown in Fig.~\ref{fig:eapd_simple}. This means that the average denoted by the bracket in \eqref{EAPD} was evaluated on the ensemble representing this torus. One can clearly see the sudden increase in distance starting from $n = 1002$. The slope of the linear fit is $\lambda =0.373 \pm 0.006$. After this, the curve goes into saturation, since the available phase space is of finite volume, thus once all trajectories of the snapshot torus become chaotic, their average distance cannot grow any further. This can be thought of as the typical behavior of snapshot tori in Hamiltonian systems subjected to parameter drift, at least in cases where the dynamics tends toward stronger chaos as the consequence of the drift, which seems to be the case here. For example, in Fig.~\ref{fig:stac}c the red torus is in the middle of the island, surrounded by other KAM tori. However, by the end of the parameter drift scenario at $n = 2100$, it has dissolved into a chaotic sea that spans the entire outer region of the phase space, meaning that all of the tori that were outside the red one had to have broken up at some point.

We note that during the period when the dynamics is still regular (i.e. before $n = 1002$), the distance between point pairs does grow, although slowly. An explanation for this could be that since the assigned pairs for each point are not exactly on the snapshot torus, during the slowly deforming phase they drift slightly further apart from the torus. Also, there is a short jump at the very beginning of the curve, which likely happens because the initial assigned pairs technically correspond to different tori (since they continuously cover the island), possessing different winding numbers from the one investigated.

The chosen form of the parameter drift in \eqref{drift} representing a mass transfer makes for a nontrivial dynamics even compared to previous studies \cite{Janosi2019,Janosi2021,Janosi2022,Janosi2024}, where a simple linear drift in only one parameter was applied, typically obtaining results similar to Figs.~\ref{fig:timdep_simple} and \ref{fig:eapd_simple}. In our case however, we found that the inclusion of just one more time-dependent parameter, even with the same drift rate, leads to more diverse dynamics on a broad scale of parameters $H$ and $L_z$. We provide an example of such peculiar dynamics in Fig.~\ref{fig:timdep_intermediate} by following the snapshot torus initiated from the red KAM torus in Fig.~\ref{fig:stac}b. Another example is shown in Figs.~\ref{fig:timdep_complex} and \ref{fig:eapd_complex} of the Appendix.

The first significant part of the evolution of the snapshot torus in Fig.~\ref{fig:timdep_intermediate} begins at $n = 50$ with its corners extending until about $p_R = \pm 350$, while some trajectories remain on the previous border of the torus, essentially closing in parts of the phase space. A similar thing happens around $p_R = 0$ as well. The created three "closed curves" then seem to be evolving on their own (panels c-f), inside the outer points of the snapshot torus, until about $n = 100$, when they experience stretching, as if they were separate snapshot tori themselves, and start to dissolve. By around $n = 482$ these points have fully scattered into a snapshot chaotic sea.
Another peculiar feature to mention is the appearance of small crinkles at $n = 255$, developing on the left side of the torus, the shape of which later dissolve, creating a narrow band (panel h).

The behavior of the snapshot torus between $n = 50$ and $n = 140$ can be explained if one takes a look at the initial phase space of Fig.~\ref{fig:stac}b. Outside of the red torus, about $p_R = \pm 500$ there are two smaller islands. Within these islands there are KAM tori just like in the largest one, and if we were to choose these tori as initial ensembles, they would become snapshot tori as well. Along those two islands we may also see four even smaller elongated islands scattered in the chaotic sea at around $p_R = \pm 200 ,\; \pm 350$. In the time-dependent scenario all those islands move and deform in the phase space, approaching the outside of the snapshot torus in Fig.~\ref{fig:timdep_intermediate}. However, what ends up happening is that points of the large snapshot torus {\em wrap around} all the smaller islands. This state holds until $n = 100$ when the snapshot tori within the small islands start to breakup, taking with them the points of the large snapshot torus around them. In the same way, the behavior around $n = 255$ is the result of its points wrapping around an island chain consisting of even more smaller snapshot tori.

The EAPD curve for this snapshot torus is shown in Fig.~\ref{fig:eapd_intermediate}, and in this case {\em there is no clear section of exponential divergence} of the trajectories, (like the one we saw in Fig.~\ref{fig:eapd_simple}). The jump in the beginning can again be explained by the initial difference in winding numbers, but for the rest of the curve the distance only grows slowly, never reaching above $\rho = -10$. This is despite the fact that by the end of the simulation ($n = 482$) the trajectories are spread out in an extended region of the phase space. 
This means that the dynamics of this snapshot torus is never (fully) chaotic, there has to be some other mechanisms at play. 

Let us remember that in time-independent Hamiltonian phase spaces on the border of islands there exist the so-called sticky regions, which exhibit non-chaotic dynamics, outside of which one finds the chaotic sea. For systems subjected to parameter drift, it was found \citep{Janosi2024}

that the picture is even more intricate. It turns out, that even when one chooses a stationary chaotic sea as initial ensemble, when it is evolved in time as a snapshot chaotic sea, some regions in it could exhibit non-chaotic dynamics. These regions are termed time-dependent non-chaotic regions, and trajectories in them can either become chaotic quickly, or stay non-chaotic for longer times. The subsets of time-dependent non-chaotic regions where the slow dynamics is persistent are called time-dependent non-hyperbolic regions, and such latter sets are thought to be the time-dependent analogs of sticky regions. Since on the EAPD curve of Fig.~\ref{fig:eapd_intermediate} no chaotic section was found, and because in Fig.~\ref{fig:timdep_intermediate} points of the snapshot torus was observed to wrap around islands of other snapshot tori, we can speculate that the time-dependent non-hyperbolic regions described above, existing around the smaller islands, might be responsible for slowing down the dynamics of the large snapshot torus.

\subsection{Time-dependent energy}
\label{sec:energy}

Changing the parameter in a Hamiltonian system results in a violation of energy conservation. The dynamics of the system can be described by the well-known adiabatic invariant for sufficiently slow parameter shifts, which is represented by the volume enclosed by a periodic trajectory in phase space. In this framework, instead of total energy conservation, an adiabatic invariant can be considered a constant of motion, provided that the time scale of the periodic orbit is much smaller than that of the parameter change. The conservation of the action variable depends on the slowness of the parameter variation \citep{Lichtenberg1992}.

\begin{figure}[!htb]
    \centering
    \includegraphics[width=\linewidth]{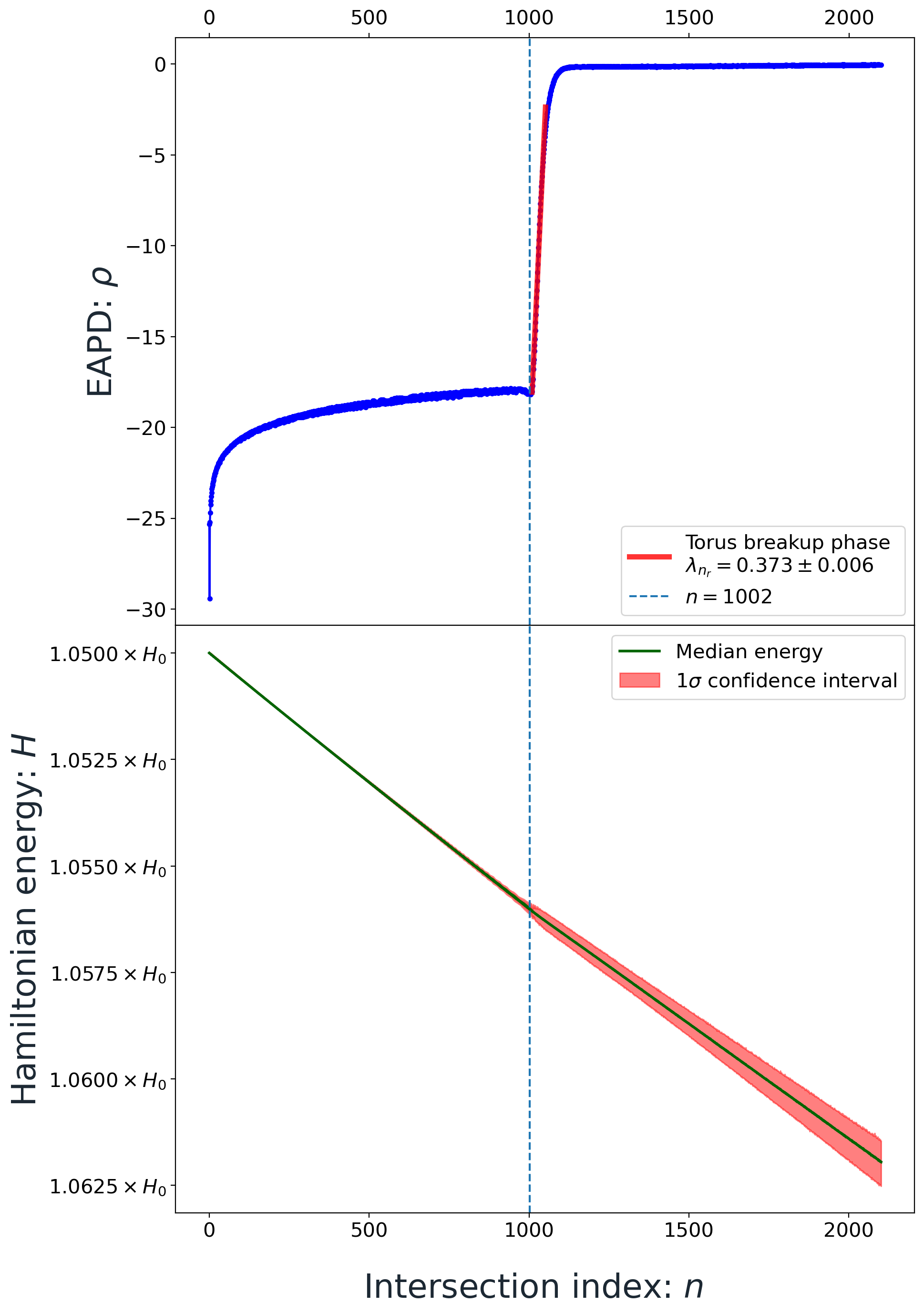}
    \caption{Median energy of the ensemble of the snapshot torus shown in Fig.~\ref{fig:timdep_simple} compared to the EAPD curve of Fig.~\ref{fig:eapd_simple}. The $1\sigma$ dispersion is shown as a light red band around the energy curve, while the vertical dashed line marks $n = 1002$, the start of the torus breakup.
    %N HELYETT KIS n
    }
    \label{fig:energy_simple}
\end{figure}

\begin{figure}[!htb]
    \centering
    \includegraphics[width=\linewidth]{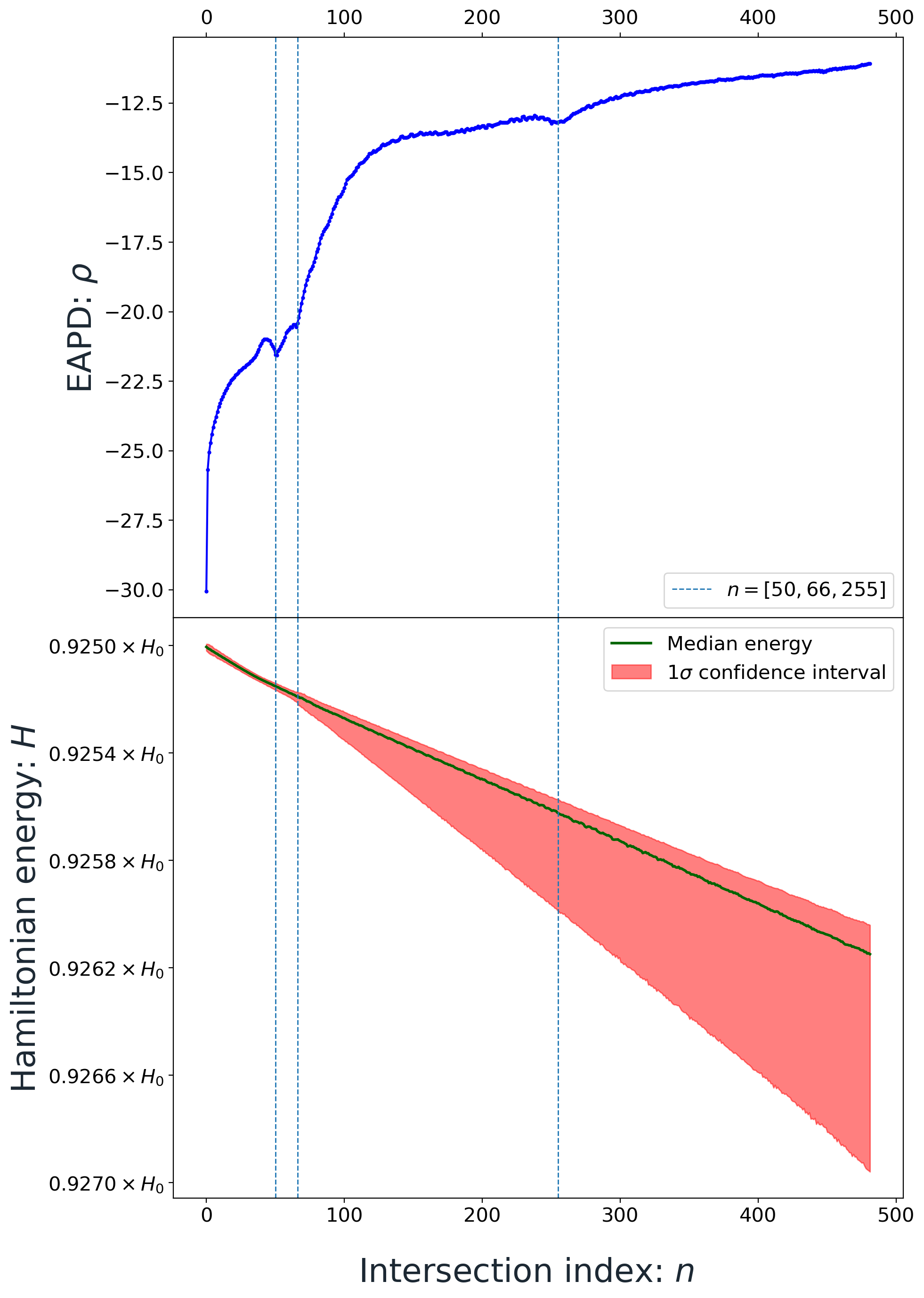}
    \caption{Median energy of the ensemble of the snapshot torus shown in Fig.~\ref{fig:timdep_intermediate} compared to the EAPD curve of Fig.~\ref{fig:eapd_intermediate}. The $1\sigma$ dispersion is shown as a light red band around the energy curve, while the vertical dashed lines mark $n = 50, 66, 255$, during the process of the snapshot torus wrapping around smaller islands.
    }
    \label{fig:energy_intermediate}
\end{figure}

In this study, the speed of parameter variation is chosen such that the system evolves beyond the adiabatic limit. That is, we do not expect even the action variable to remain constant during the motion. This fact is also evident from the performed numerical simulations, where not only the geometric structure but also the size of KAM islands changes over time in phase portraits. During the mass exchange between the two parts of the galaxy, Eq.~(\ref{drift}), the total energy of the system is monitored. It turns out that - see Figures \ref{fig:energy_simple} and \ref{fig:energy_intermediate} - the system becomes increasingly bound as the median energy decreases. The energy curves are plotted alongside the respective EAPD curves of Figs.~\ref{fig:eapd_simple} and \ref{fig:eapd_intermediate}, in order to compare and highlight relevant parts of the processes. The relative change in total energy averaged over the ensemble is less than 1.2\% by the end of each integration. In Fig.~\ref{fig:energy_simple}, along the energy curve of the snapshot torus shown in Fig.~\ref{fig:timdep_simple}, a slight break is visible at the time of torus breakup. This characteristic becomes even more pronounced when considering the energy dispersion ($\sigma$) evaluated over the ensemble. The sudden change in $\sigma$ around the disappearance of the torus can be regarded as an alternative indicator of the dramatic structural evolution of KAM islands in phase space during parameter changes.

For the case of the snapshot torus belonging to Fig.~\ref{fig:timdep_intermediate}, 
the energy also monotonically decreases, with the dispersion growing much more widely than in Fig.\ref{fig:energy_simple}, starting at around $n = 50$.

The Hamiltonian function (\ref{Ham}) determines conservative dynamics while the parameters remain unchanged over time. However, if we allow the mass between the disk and the bulge to evolve, the internal dynamics experience time-dependent perturbation, which affects the initial energy of the system, as shown here. In other words, manually setting the parameters appears as an external disturbance in the dynamics. Consequently, we model stellar dynamics in a time-dependent potential without employing any concrete physical mechanisms. However, similar results are expected if we had taken into account an additional term explicitly in the effective potential Eq.~(\ref{allpot}) controlling mass transfer within the galaxy.

\section{Conclusions}
\label{sec:4}

In this paper we have numerically studied ensembles of stellar obits for the Milky Way with different potential profiles for the galactic components: the extended Myamoto-Nagai potential for the disk, the Hernquist potential for the bulge, and the NFW potential for the dark matter halo. We simulated a mass transfer from the disk to the bulge by heuristically introducing a decreasing and increasing parameter drift of the same rate in the mass parameters of the disk and the bulge, respectively. 
%he observational basis for this that during the evolution of spiral galaxies mass is found to traverse from the disk to the bulge, as the galaxy transforms into an elliptic one REFS HERE 
For simplicity, for the form of the drift we chose a linear function; however, in general cases, any monotonous form could be used obtained from the observations. 
{The chosen rate results in 30\% of the disk mass transferred to the bulge on timescales that can be deemed reasonably long compared to the timescales of our Universe \citep{Spergel1997}. }

The stellar orbits are illustrated on Poincaré SoSs in the case of constant mass, and they are shown to exhibit a mixed phase space with coexisting periodic, quasi-periodic, chaotic and sticky dynamics.
When the parameter drift is introduced, as an appropriate ensemble of initial conditions we chose quasi-periodic KAM tori from the stationary case. 
In the time-dependent dynamics they become snapshot tori, which change their shape in the phase space and in one of the investigated cases their dynamics becomes chaotic through a breakup process.

The torus breakup results in an exponential separation of nearby points on the torus. This is characterized by the so-called EAPD curve, which describes the evolution of the average distance of point pairs. The slope of this curve can be thought of as a new type of finite-time Lyapunov exponent that characterizes the fate of the torus. In the case mentioned above, this Lyapunov exponent is approximately constant and strictly positive through the breakup process. However, we found that in other cases the EAPD curve does not contain any exponential parts, the dynamics of the torus never turns chaotic. This slow dynamics results in nontrivial Lyapunov exponents throughout the whole process. We note that the two presented cases are far from fully covering the rich dynamics of the problem, and further studies are needed for its comprehensive discovery.

As is noted in Sec.~\ref{sec:energy}, the snapshot pictures comprised of different intersections with various time instants of the parameter drift can result in changing trajectory energies. Future studies will be necessary to further explore the emerging changes induced by different parameter changes in order to better understand the dynamics of time-dependent systems. The change in trajectory energies may play a significant role in inducing local, or even global, changes in the morphology of these galaxy models, such as the emergence of more dense star-forming regions with distinct structures by purely dynamical causes (\cite{fractal_struct,spiral,spatial_auto}).
Classical bulges are believed to form through tidal disruption events and galactic mergers \citep{mergers}, which are violent processes taking place on relatively short timescales. In contrast, pseudo-bulges are believed to form and evolve through secular processes \citep{2004ARA&A..42..603K, bulge&pseudo}, contributing to the slow buildup and reshaping of pseudo-bulges, in which continuous mass transfer between galactic components may also play a role.

The general aim of this paper is to provide, through the presented example, a universal tool to describe the problems arising in galactic and celestial mechanics where a parameter drift is present. The method originates in chaos theory, and is currently an actively developed area of research. Some other concepts that could be useful from this method include time-dependent stable and unstable manifolds and foliations \citep{Janosi2024}, doubly transient chaos \citep{Motter2013}, as well as snapshot attractors in dissipative cases \citep{Janosi2019, JKT21, Janosi2022,Janosi2024}.
Some further methods that could be used for augmenting the presented methods include, for example, developing new ways of phase space projections dependent on trajectory energies, or by expanding the EAPD analysis with mapping methods.
Furthermore, it can also be beneficial to expand the analysis into more composite and realistic N-body simulation models, where the axisymmetric bulge and dark matter halo components and the non-axisymmetric spiral arms are also included in the potential (\cite{Barros2016, Ektoras2016,Wilma2017,Michtchenko2018}).

\begin{acknowledgements}
{We are grateful to Tamás Tél and Santi Roca-Fàbrega for their valuable comments, which significantly improved the interpretation of the results.}
This work was supported by the Hungarian National Research, Development and Innovation Office, under Grants No.~KDP-2023 C2262591,  TKP--2021 BME-NVA-02 (D.J.) and TKP2021-NKTA-64 (T.K.), financed by the Ministry of Culture and Innovation of Hungary.

\end{acknowledgements}

   \bibliographystyle{aa} % style aa.bst
   \bibliography{references} % your references Yourfile.bib

\begin{appendix}
\label{appendix}

\section{Illustrating another example of peculiar dynamics}
\begin{figure}[!htb]
    \centering
    \includegraphics[width=\linewidth]{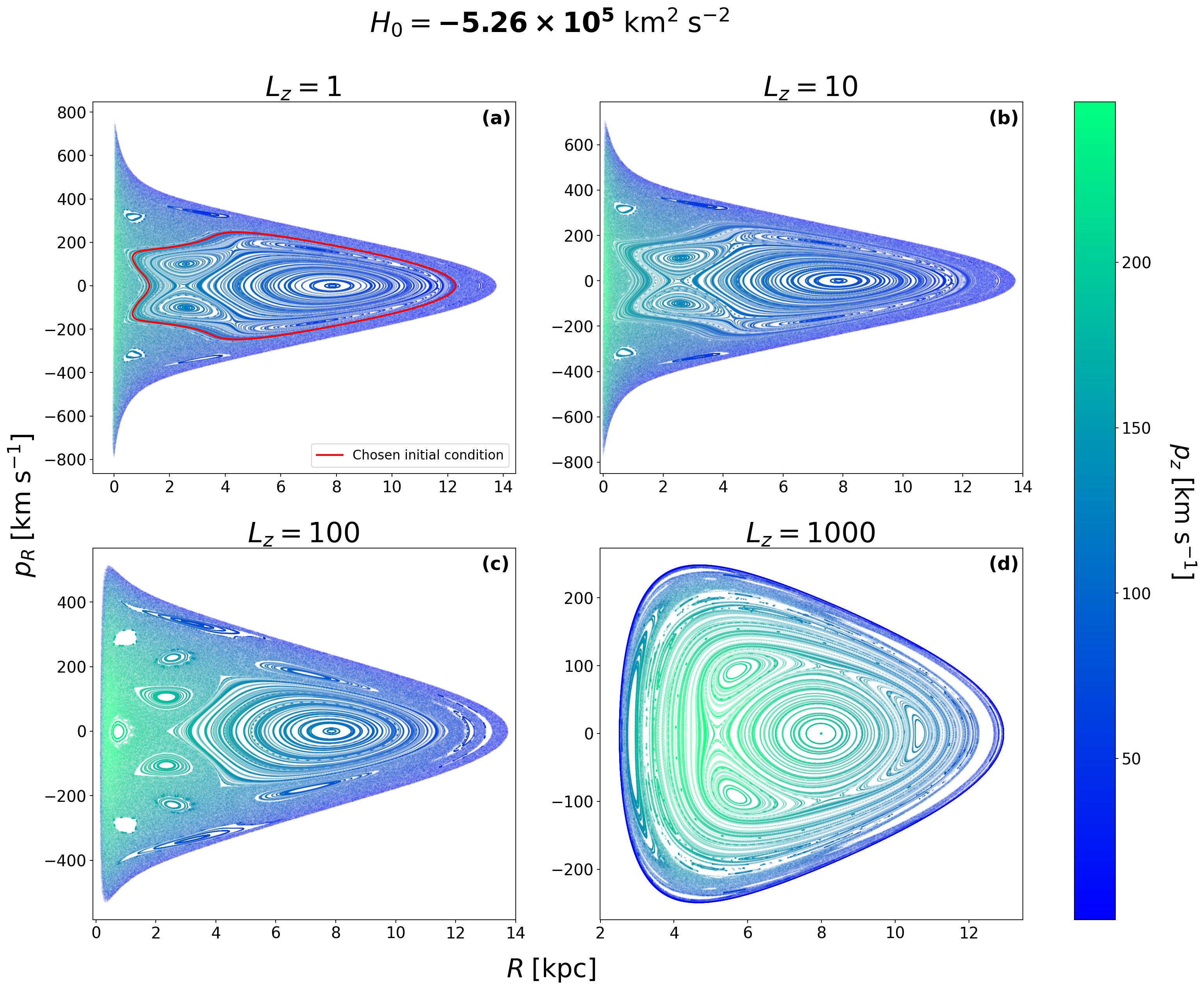}
    \caption{Phase portraits of Eq.~\eqref{eqm} for different values of $L_z$ with constant $H=H_0$. The trajectories are plotted on the $(R,p_R)$ plane, following the Poincaré cut at $z = 0$, and the value of $p_z$ obtained from \eqref{Ham}, indicated at each point by the colormap. The red KAM torus belongs to the initial condition $R_0 =0.67, \; {p_R}_0 =  129.79$ (panel a).} 
    \label{fig:stac2}
\end{figure}

Figure~\ref{fig:stac2} shows the phase space plots in the $(R,p_R)$ plane, where we illustrate the phase portrait for four different values of $L_z$ with the calculated $H_0$ value. 

\begin{figure}[!htb]
    \centering
    \includegraphics[width=\linewidth]{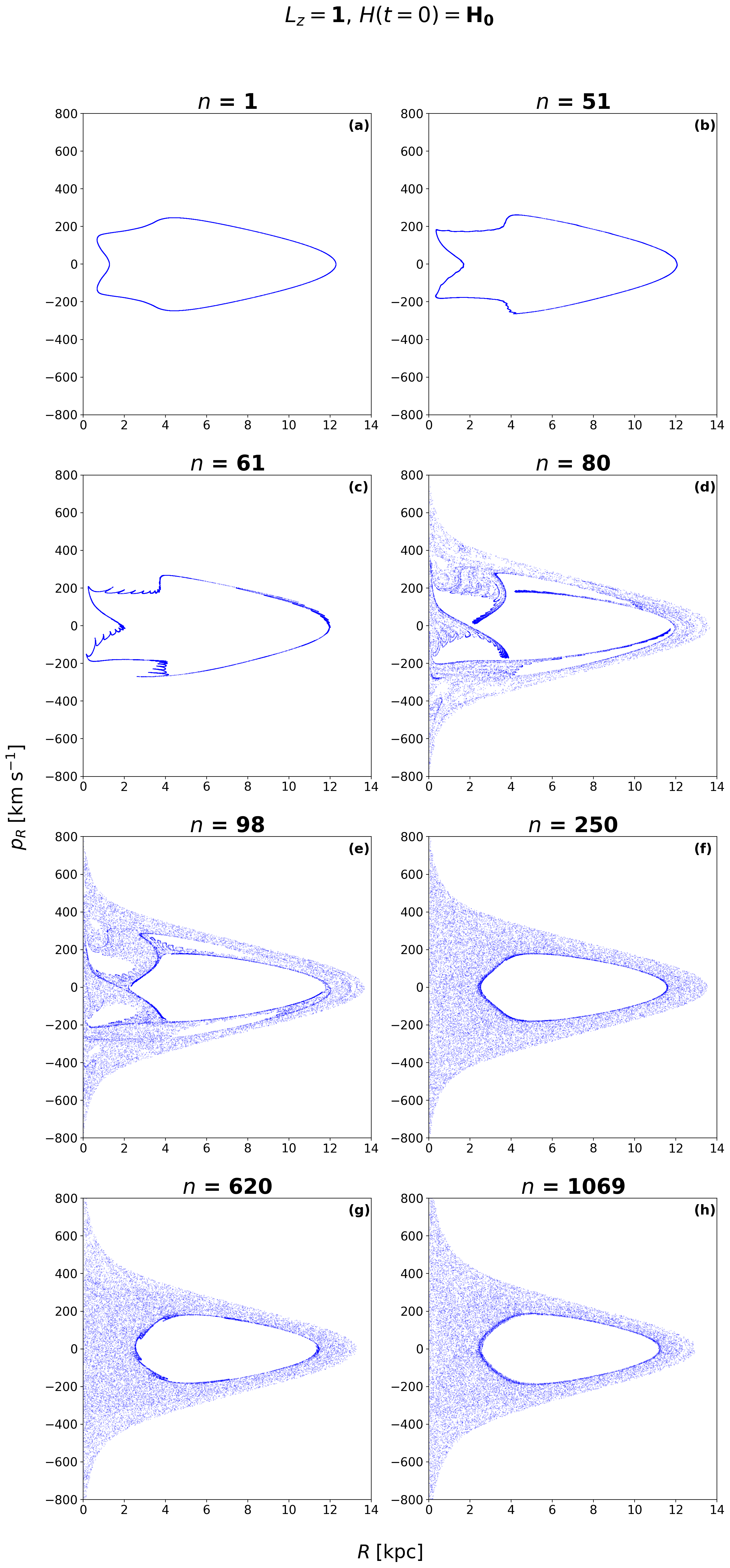}
    \caption{Time evolution of the red torus on Fig.~\ref{fig:stac2}a. At $n = 61$ we can see the initial torus starts to deform significantly, while a small fraction of trajectories scatter out into the chaotic sea after $n = 80$. After $n=250$ 
    only an inner structure remains, still separating snapshot tori from the snapshot chaotic sea.
    }
    \label{fig:timdep_complex}
\end{figure}

A case of even more peculiar dynamics is shown in Fig.~\ref{fig:timdep_complex} by following the snapshot torus shown in red in Fig.~\ref{fig:stac2}a, similarly to the previous two cases in the main text. In this case an ensemble of orbits is chosen which have significantly lower angular momentum $L_z$, hence also having significantly low radial velocities compared to the cases presented before. 
The first stage of its evolution is similar to the torus presented in Fig.~\ref{fig:timdep_simple}, where the torus develops stretches, 
which results in its deformation. However, the torus does not breakup the same way as the one in Fig.~\ref{fig:timdep_simple} does, 
this time the deformation begins from 
several small tongues on the left side of the torus at $n = 61$. These then transport points into the snapshot chaotic sea, but, at the same time, four tongues can be observed to penetrate into the torus and wrapping around areas that were originally {\em inside} it, see instants $n = 80, 98$.
These structures are similar to the islands seen in Fig.~\ref{fig:timdep_intermediate}, disappearing by $n = 250$.
However, even at this point a significant part of the snapshot torus remains intact, with a large number of points from the original structure still bordering this new shape, similarly to the case of Fig.~\ref{fig:timdep_intermediate}. This property remains true until the end of the simulation at $n = 1069$.
The behavior of the peculiar structures penetrating inside the large island, present in the snapshots of $n=80,98$, are not yet understood and further, more extended research is needed.

\begin{figure}[!htb]
    \centering
    \includegraphics[width=\linewidth]{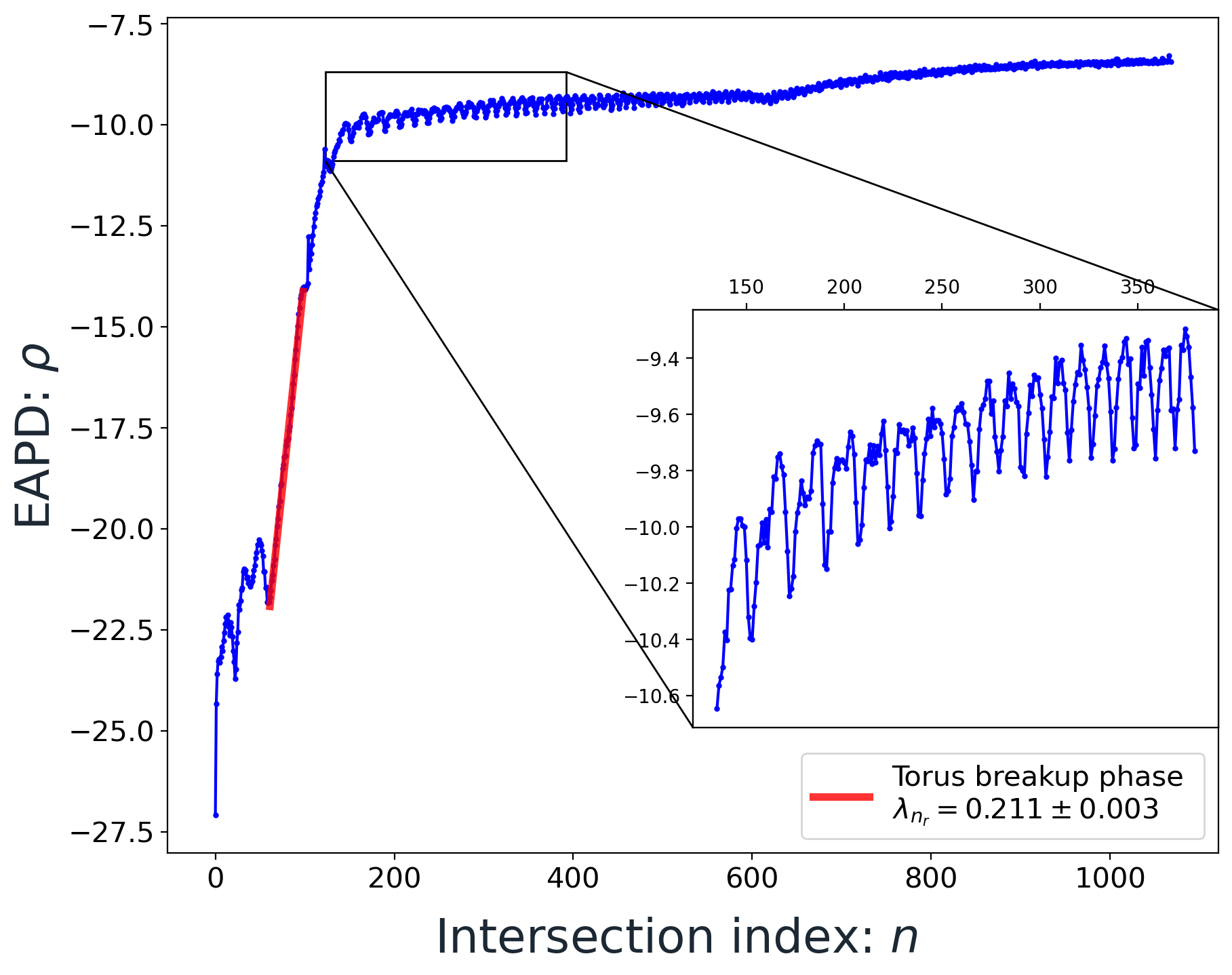}
    \caption{EAPD curve of Eq.~\eqref{EAPD} evaluated for the snapshot torus of Fig.~\ref{fig:timdep_complex}. The inset shows an oscillating behavior of small amplitudes after $n = 150$. 
    A breakup phase is present after $n=61$, for which the Lyapunov exponent obtained from the fitting is $\lambda_{n_r} = 0.211 \pm 0.003$. }
    \label{fig:eapd_complex}
\end{figure}

The EAPD curve for this torus is shown in Fig.~\ref{fig:eapd_complex}, where we see an initial exponential increase from $n = 61$, but after the points finished scattering into the outer chaotic sea the distance goes into saturation at a value orders of magnitude smaller than unity, consistent with the observations of Fig.~\ref{fig:timdep_complex} at $n = 620, 1069$.

\begin{figure}[!htb]
    \centering
    \includegraphics[width=\linewidth]{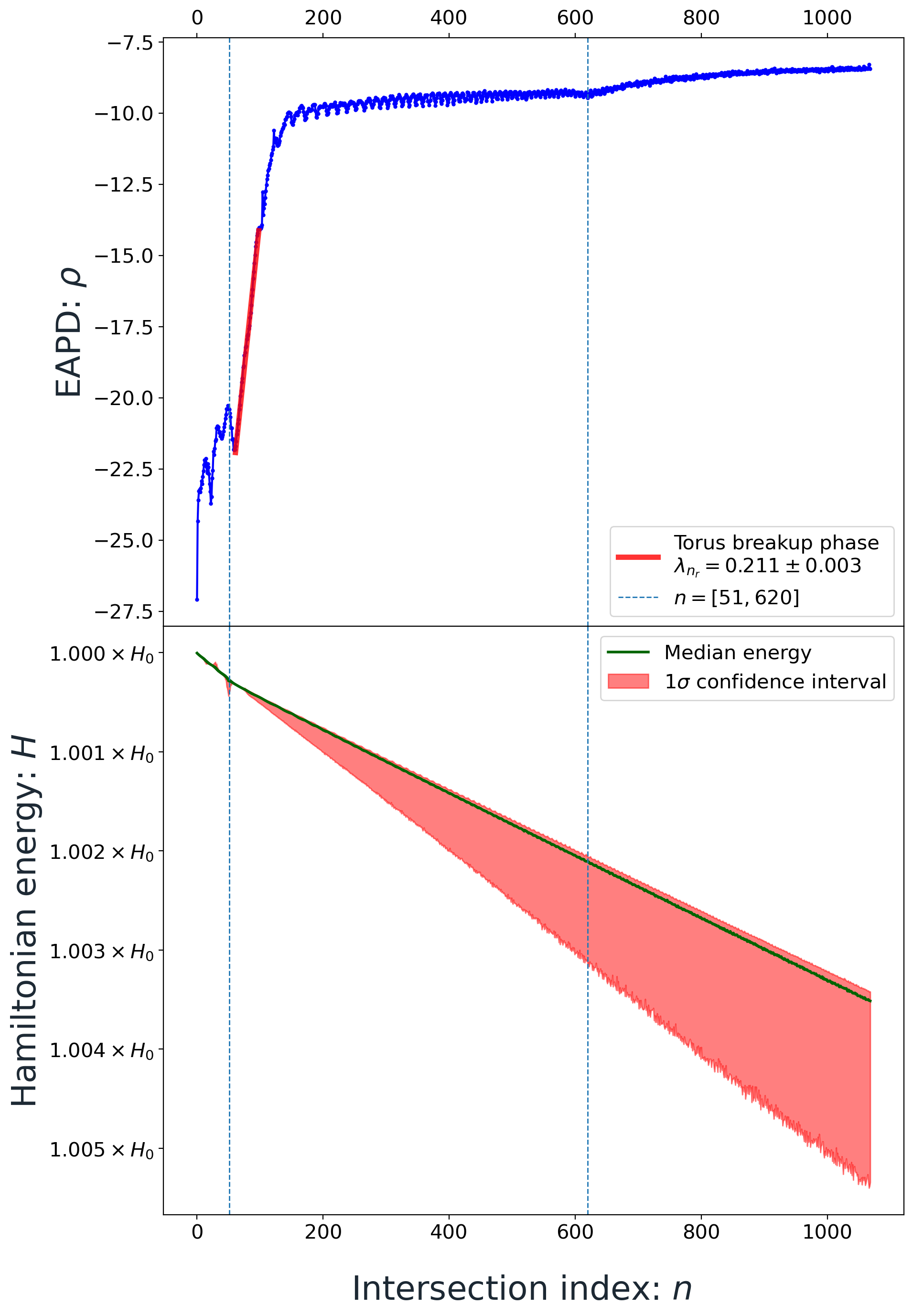}
    \caption{Median energy of the ensemble of the snapshot torus shown in Fig. \ref{fig:timdep_complex} compared to the EAPD curve of Fig.~\ref{fig:eapd_complex}. The $1\sigma$ dispersion is shown as a light red band around the energy curve, the vertical dashed lines mark $n = 51, 620$. 
    }
    \label{fig:energy_complex}
\end{figure}

In Fig.~\ref{fig:energy_complex}, a monotonous growth in energy dispersion can be seen, similarly to the case in Fig.~\ref{fig:timdep_intermediate}, and after $n = 51$ the mean curve in energy shows a slight break.

\end{appendix}
\end{document}